\documentclass[preprint, 12pt]{elsarticle} 

\usepackage{lineno}
\usepackage{subfigure}

\usepackage{bm}
\usepackage{amsmath,amssymb,mathrsfs}
\usepackage{latexsym}
\usepackage{graphicx} 
\usepackage{epstopdf}
\usepackage{graphicx,epstopdf,color}
\usepackage{amsfonts}
\usepackage{bm}
\usepackage{bbm}
\usepackage{multirow}

\journal{arXiv}

\begin{document}

\begin{frontmatter}


\title{Performance Comparison of UCA and UCCA based Real-time Sound Source Localization Systems using Circular Harmonics SRP Method}



\author[label1,label2]{Zhe Zhang}
\author[label1,label2]{Ming Wu}
\author[label1,label2]{Xinyu Han}
\author[label1,label2]{Jun Yang}

\address[label1]{Institute of Acoustics, Chinese Academy of Sciences, No. 21 North 4th Ring Road, Haidian District, 100190 Beijing, China}
\address[label2]{University of Chinese Academy of Sciences, No.19(A) Yuquan Road, Shijingshan District, 100049 Beijing, China}

\begin{abstract}
Many sound source localization (SSL) algorithms based on circular microphone array (CMA), including uniform circular array (UCA) and uniform concentric circular array (UCCA), have been well developed and verified via computer simulations and offline processing. On the other hand, beamforming in the harmonic domain has been shown to be a very efficient tool for broadband sound source localization due to its frequency-invariant properties. In this paper, design, implementation, and performance of a real-time sound source localization system are discussed. Specifically, we analyze the effect of parameter settings and compare the performance between UCA and UCCA through real-time experiments in a real room. The proposed method shows significant improvement by using UCCA instead of UCA.
\end{abstract}


\end{frontmatter}


\section{Introduction}
\label{S:1}

Sound source localization based on microphone arrays has applications in many areas such as immersive audio playback, artificial audition, and human-computer interaction \cite{escudero_real-time_2018, sun_localization_2012, valin_robust_2003}. In recent years, different kinds of algorithms have been developed to deal with broadband sound source localization attempting to improve accuracy and robustness \cite{chen_source_2002} \cite{meyer_spherical_2008, teutsch_acoustic_2006}. Methods based on circular harmonics beamforming (CHB), or referred to as eigenbeamforming, have shown to provide better localization performance in resolution and sidelobe while reducing computational load due to its frequency-invariant properties \cite{torres_robust_2012, belloni_beamspace_2006}. Meanwhile, approaches based on uniform concentric circular array (UCCA) \cite{chan_uniform_2007} instead of traditional uniform circular array (UCA) \cite{tiana-roig_beamforming_2010} are proposed to achieve better performance. There are many works focused on the theoretical performance of beamformers and offline processing analysis \cite{parthy_comparison_2011, ishi_evaluation_2009} . However, only a few describe the actual implementation of real-time sound source localization systems and its use in real environments with noise and reverberation \cite{rabinkin_dsp_1996, seunghun_jin_real-time_2008, pavlidi_real-time_2013}. 

In the presented paper, we discuss about the design, implementation, and performance of a real-time sound source localization system using circular harmonics SRP method. The prototype system is tested in a real room with background noise and reverberation under different parameter settings to evaluate the performance. Specifically, a comparison is conducted between UCA and UCCA, which shows a significant improvement in accuracy and robustness when using UCCA.

This paper is organized as follows. In section 2, we give a brief review on the proposed method and its optimization. In section 3, descriptions about the hardware design and software development are shown. In section 4, analysis and evaluation of experiment results are presented and discussed. Section 5 concludes the paper.

\section{Proposed Method}
\label{S:2}

\subsection{Circular Harmonics Decomposition}

Consider a UCCA with $P$ rings, each having $N_p$ elements at equidistant locations on a circle of radius $r_p$, with the center of the array located at the origin of coordinates. Assuming a plane wave coming from the median plane, the sound pressure at the $p_{th}$ ring can be written in polar coordinates as

\begin{equation}
\label{eq:xpoftheta}
x_{p}\left(k r_{p}, \theta\right)=S(\omega) e^{j k r_{p} \cos \left(\theta-\theta_{i}\right)},
\end{equation}
where $S(\omega)$ is the amplitude of the impinging wave in angular frequency $\omega$. 
The above expression can be expanded into a series of circular waves and expressed as

\begin{equation}
x_{p}\left(k r_{p}, \theta\right)=S(\omega) \sum_{l=-\infty}^{\infty} j^{l} J_{l}\left(k r_{p} \right) e^{-j l \theta_{i}} e^{j l \theta},
\end{equation}
where $J_p$ is a Bessel function of the first kind of order $l$. We can define the circular harmonics decomposition coefficients as

\begin{equation}
C_{l}\left(k r_{p}, \theta_{i}, \varphi_{i}\right)=S(\omega) j^{l} J_{l}\left(k r_{p} \right) e^{-j l \theta_{i}},
\end{equation}
which can be also obtained by conducting a Fourier transform on the sound pressure $x_p$. Thus, we derived the circular harmonics composition of the sound pressure on the $p_{th}$ ring

\begin{equation}
x_{p}\left(k r_{p}, \theta\right) = \sum_{l=-\infty}^{\infty} C_{l}\left(k r_{p}, \theta_{i} \right) e^{j l \theta}.
\end{equation}

It should be noted that we have to use the infinite number of Fourier terms to present the sound pressure in practice. The relation between the number of sampling microphones $M$ and the maximum circular harmonics decomposing order $L$ is given by \cite{torres_robust_2012}

\begin{equation}
\label{eq:order}
    M \ge 2L + 1.
\end{equation}

\subsection{Modal Beamforming}

Traditional beamformer in the element domain is frequency-dependent, so it will bring a big computational burden to iterate all frequency bins interested. Taking the advantages of circular harmonics beamformer (CHB), it only needs to combine a few sets of circular harmonic components to derive the response, which is of importance for implementation in real-time. 

The compensating filter of the corresponding order can be derived from the ideal response of the beamformer:

\begin{equation}
    \sum_{l=-L}^{L} \sum_{p=1}^{P}\left\{C_{l}\left(k r_{p}, \theta_{i} \right) H_{l}^{p}\left(\omega \right)\right\} e^{j l \theta}=S(\omega) \delta\left(\theta-\theta_{i}\right),
\end{equation}
where $H_{l}^{p}\left(\omega \right)$ is the so-called compensating filter for the corresponding ring and order.

When $p=1$, it turns out to be an UCA case, and the compensating filter is solved as

\begin{equation}
    H_{l}\left(\omega\right)=\left(j^{l} J_{l}\left(k r \right)\right)^{-1}.
\end{equation}

Besides the low magnitude of high orders, it is known that the Bessel functions have periodic zeros which cause the seriously amplifying of the noise. To avoid this, the use of Tikhonove-regularization is proposed \cite{parthy_comparison_2011} as
\begin{equation}
    \bar{H}_{l}\left(\omega\right)=\frac{(-j)^{l} J_{l}\left(k r\right)}{\left\|J_{l}\left(k r \right)\right\|^{2}+\alpha},
\end{equation}
which improves the robustness of the beamformer but causes the broader beam pattern.

In circumstances of UCCA, we can derive the compensating filters by solving the optimization problem with the least square norm condition:

\begin{equation}
    \min \left\| 
    \begin{array}{cccc}
    H_{l}^{1}\left(\omega \right)\\
    H_{l}^{2}\left(\omega \right)\\
    \vdots\\
    H_{l}^{p}\left(\omega \right)
    \end{array}
    \right\|^{2} \quad \text { s.t. } 
    \left[ \begin{array}{cccc}
    j^{l} J_{l}\left(k r_{1} \right)\\
    j^{l} J_{l}\left(k r_{2} \right)\\
    \vdots\\
    j^{l} J_{l}\left(k r_{p} \right)
    \end{array} \right] ^H
    \left[ \begin{array}{cccc}
    H_{l}^{1}\left(\omega \right)\\
    H_{l}^{2}\left(\omega \right)\\
    \vdots\\
    H_{l}^{p}\left(\omega \right)
    \end{array} \right] = 1.
\end{equation}

By using the vector-form compensating filters from above, the zero-points problem of UCA is solved because the zero points differ by the radii of circles. Figure 2 shows a comparison of reciprocals of the absolute values of compensating filters between a 2-ring UCCA and its individual rings as UCAs.

\begin{figure}[h]
\centering\includegraphics[width=1\linewidth]{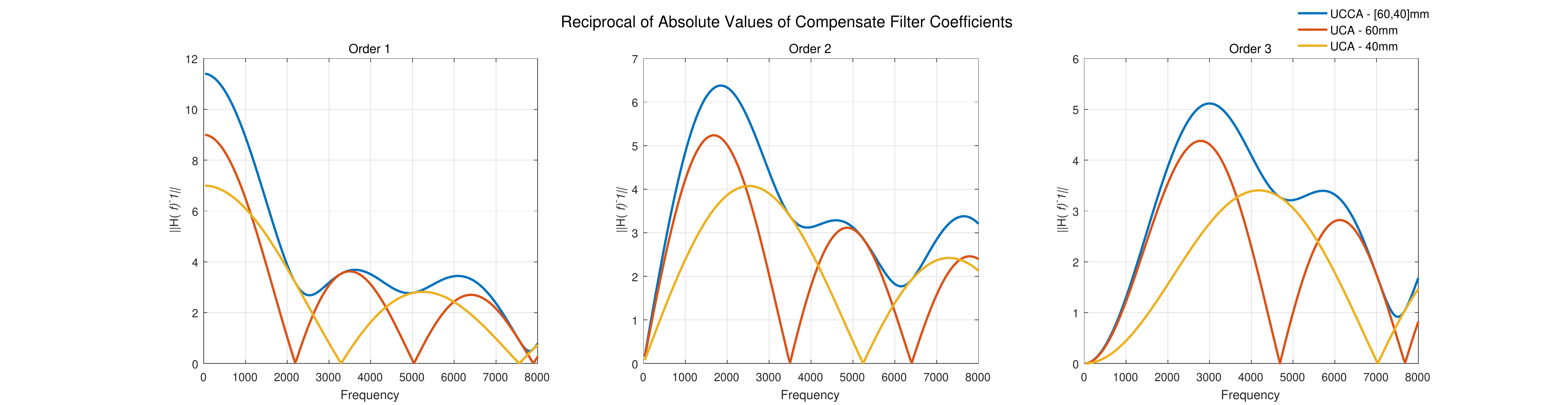}
\caption{Reciprocals of the absolute values of compensating filters for UCA and UCCA.}
\label{bessel}
\end{figure}

Once obtain the compensating filter for the microphone array, we can implement a steered-response power (SRP) approach to scan a $360^{\circ}$ spacial spectrum, given by

\begin{equation}
    P_{SRP}(\theta)=\sum_{f=f_{0}}^{f_{1}}\left[\tilde{\mathbf{C}}_{c}\left(f, \theta_{i}\right)^{T} \mathbf{A}(\theta)\right]^2,
\end{equation}
where $[f_{0},f_{1}]$ is the selected frequency range that SRP averages on, $\tilde{\mathbf{C}}_{c}$ is the compensated harmonic coefficients, and $\mathbf{A}(\theta)$ is the steering vector
\begin{equation}
    \mathbf{A}(\theta)=\left[e^{-j L \theta}, \ldots, e^{-j 0 \theta}, \ldots, e^{j L \theta}\right]^{T}.
\end{equation}

\section{Implementation}
\label{S:3}

\subsection{Hardware Design}
The block diagram of the developed prototype system and a photo of the microphone array are depicted in Figure 2. The TI TMS320C6678 DSP processor plays the central role in the real-time operations. We build a 2-Ring UCCA with 9 and 7 microphones equidistantly set on the outer ring and the inner ring, respectively. A 16 channel ADC chip transforms analog signals captured by the microphone array to digital signals and passes them to the DSP processor. The DDR3 memory in the system makes it possible to store large floating number arrays of the filter coefficients downloaded from the host PC.

\begin{figure}[h]
\centering
\includegraphics[width=0.8\linewidth]{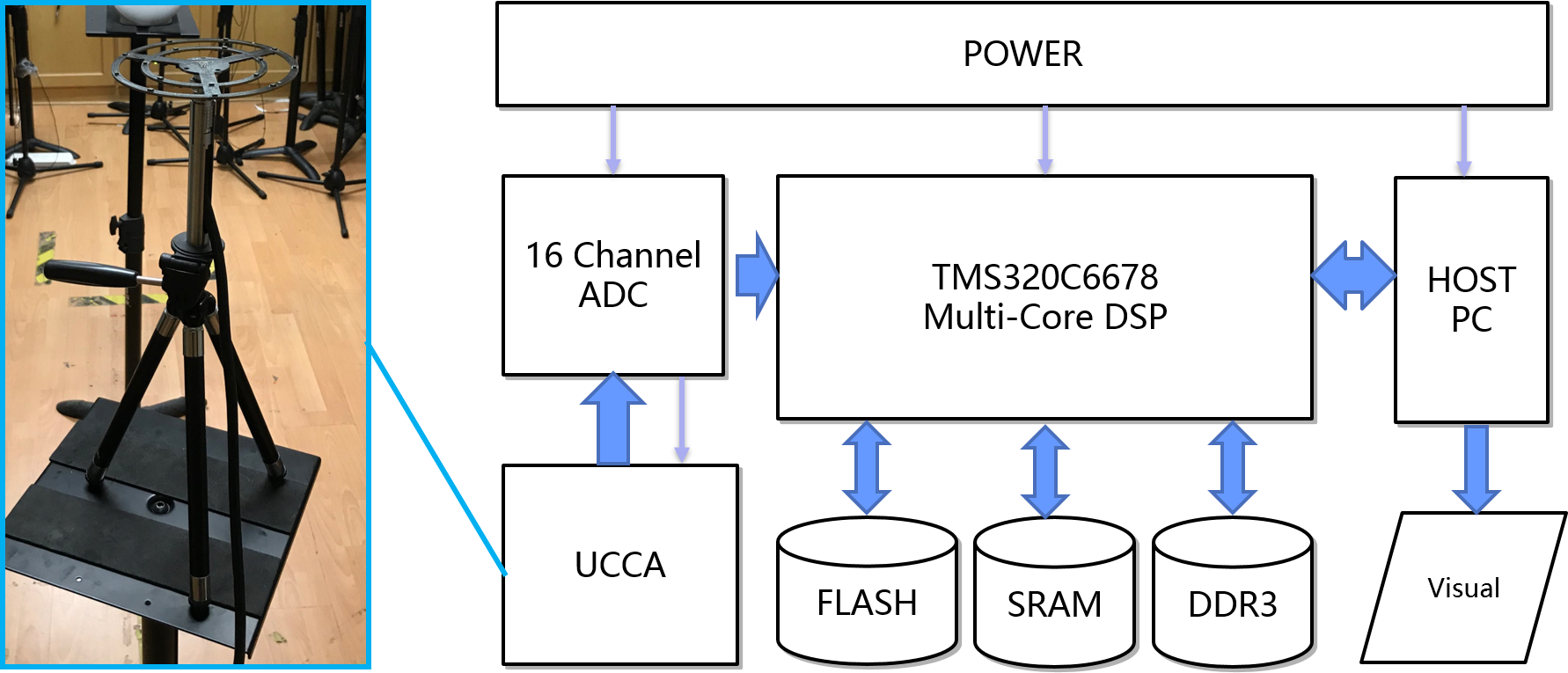}
\caption{The UCCA and block diagram of the real-time sound source visualization system.}
\end{figure}

\subsection{Software Development}
Sound is captured by the UCCA and digitized by the ADC chip at the sampling rate of 16kHz. Thanks to the high performance of the DSP processor and its KeyStone multi-core architecture, the master core plays the role of data retrieval and communications with host PC, and simultaneously the slave cores fulfills the tasks of manipulating matrices for the algorithms and return the result to the master core. The steering vector and compensating filters are initialized when the program starts to run. In the real-time block processing, a 512-sample frame of data from each channel is gathered and an FFT with Blackman windows is performed on all channels in the master core. The master core then sends the FFT data in the selected frequency range (up to 8kHz in this 16kHz-sampling case) to the slave core. While the slave core is doing the calculations of algorithms in the current frame, the master core reads and sends to host PC the calculated results of the last frame that the slave core has finished calculating. Then the loop starts over again when the timer calls the master core to execute the tasks on the next frame.

In addition, it is a bit troublesome to manipulate complex value matrices in DSP systems, with existing libraries often limiting users in certain conditions like a fixed word length. To make it easy to realize the algorithms, we start library functions from scratch, taking advantage of the structure and dynamic memories in C programming, and build a framework suitable for implementation of SSL algorithms. This helped we vividly demonstrate the proposed method based on our system.

By implementing such software scheme, we guarantee that the system works in real-time with a low latency level: one-frame delay (32ms under the sampling rate and the buffer size we use now). Anyway, there may also be extra latency caused by other procedures that should be taken into consideration in practical applications.

\section{Performance Evaluation}
\label{S:4}

\subsection{Parameters Analysis}
According to equation \ref{eq:order}, our microphone array with 9 and 7 elements on the corresponding ring has a max modal order 3. Plus, Nyquist sampling theorem determines the spatial aliasing frequency, which is given by
\begin{equation}
    f_{al} = c/2d,
\end{equation}
where $c$ is the sound speed and $d$ is the inter-element distance.

In practice, we can artificially separate the signals captured by different rings of the UCCA in our program. Thus, the UCCA system can work as a small UCA and a large UCA separately with the corresponding radius of $4cm$ and $6cm$, ignoring the scattering of the structure in the sound field.

Taking above into account, the CH-SRP algorithm described in Section 2 and the prototype system described in Section 3 were set up and tested in different parameter settings in a room with a $T_{60}=0.42s$. The parameter settings are shown in Table 1.

\begin{table}[h]
\centering
\begin{tabular}{c c c c c}
\hline
\textbf{CMA} & \textbf{Channels} & \textbf{Radius} & \textbf{Max Order} & \textbf{Freq Range(kHz)}\\
\hline
$UCA_S$ & 7 & 4cm & 1, 2, 3 & [1,2], [2,3], [3,4]\\
$UCA_L$ & 9 & 6cm & 1, 2, 3 & [1,2], [2,3], [3,4]\\
$UCCA$ & (9, 7) & (6, 4)cm & 1, 2, 3 & [1,2], [2,3], [3,4]\\
\hline
\end{tabular}
\caption{Parameter settings}
\end{table}

A YAMAHA HS 5 speaker lined with an Audient iD14 audio interface is set $2m$ from the microphone array. We play white noise for a period time while recording using the UCCA. A total length of 1024 frames data are stored and processed under different parameter settings. The SRP scanning step-size is set to $3^{\circ}$, meaning that every frame the system returns a 120-floating-number array as the estimated spectrum, in which the index of  maximum value corresponds to the azimuth estimated. The average values and standard deviations are shown in Figure 3, and the spacial spectra plotted is shown in Figure 4.

\begin{figure}[htb]
    \centering
    \includegraphics[width=1\linewidth]{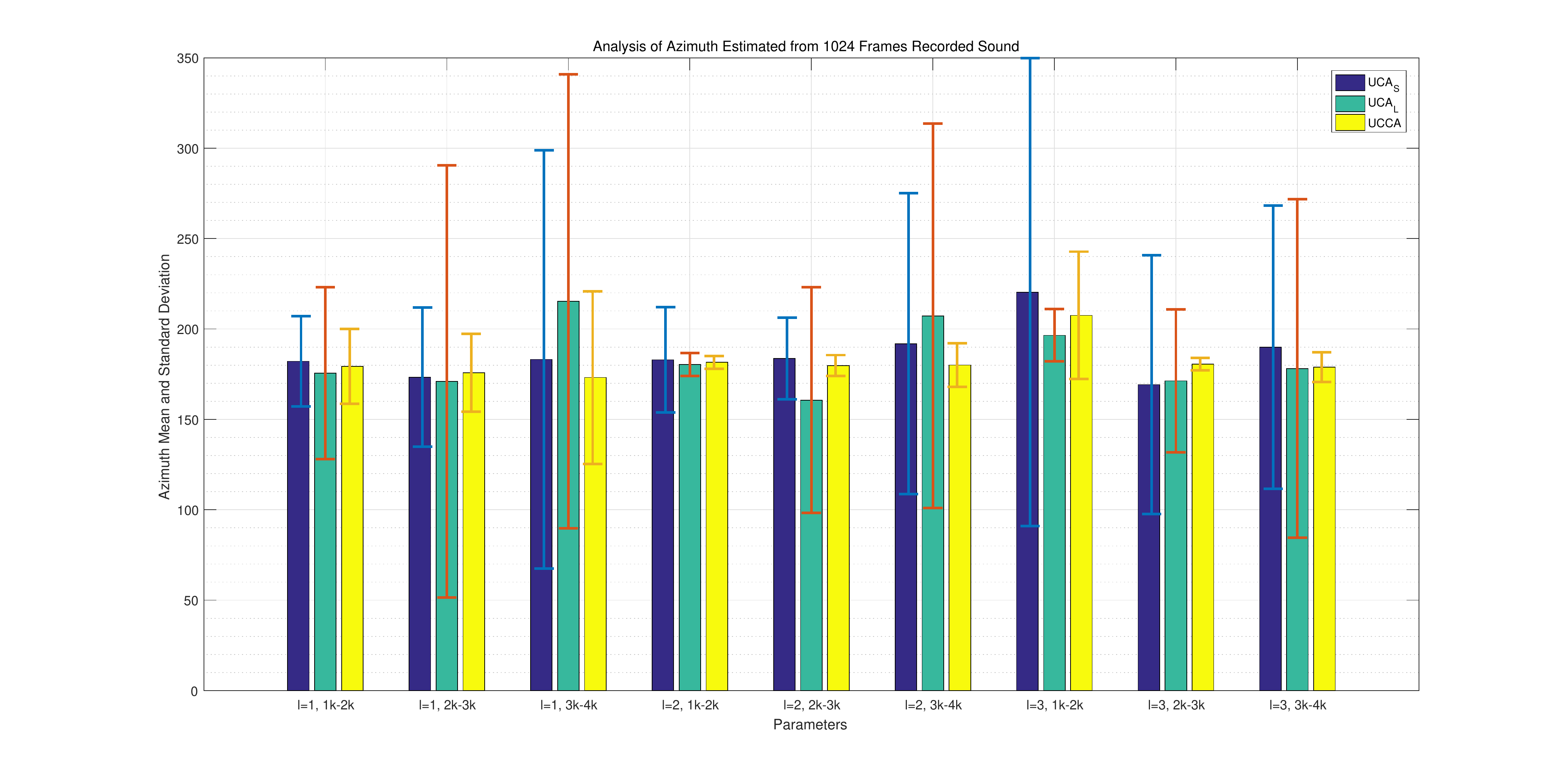}
    \caption{Mean and standard deviation of azimuth estimated from 1024 frames.}
    \label{fig:errorbar}
\end{figure}

\begin{figure}[p]
\centering\includegraphics[width=0.9\linewidth]{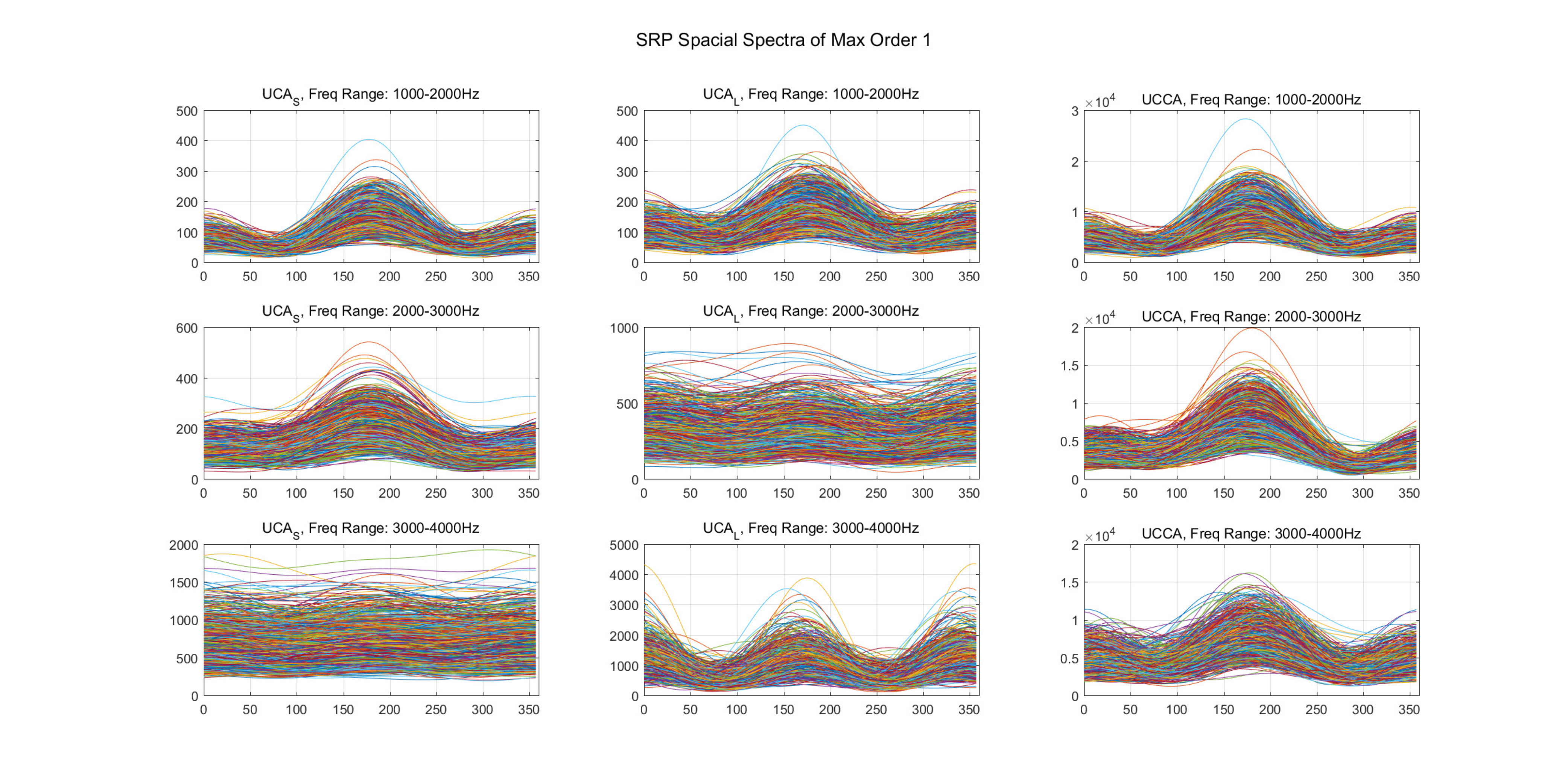} \\
\centering\includegraphics[width=0.9\linewidth]{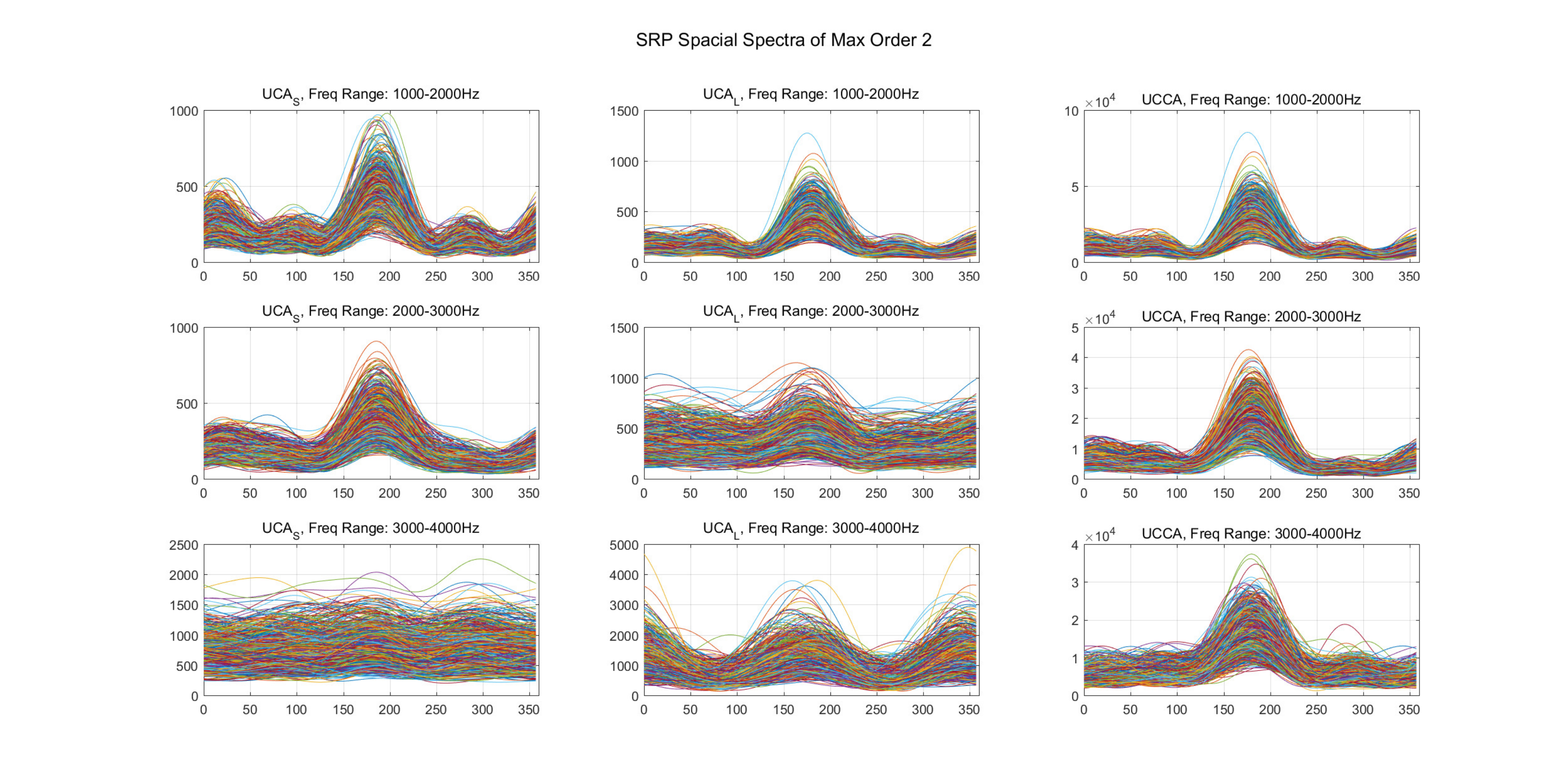} \\
\centering\includegraphics[width=0.9\linewidth]{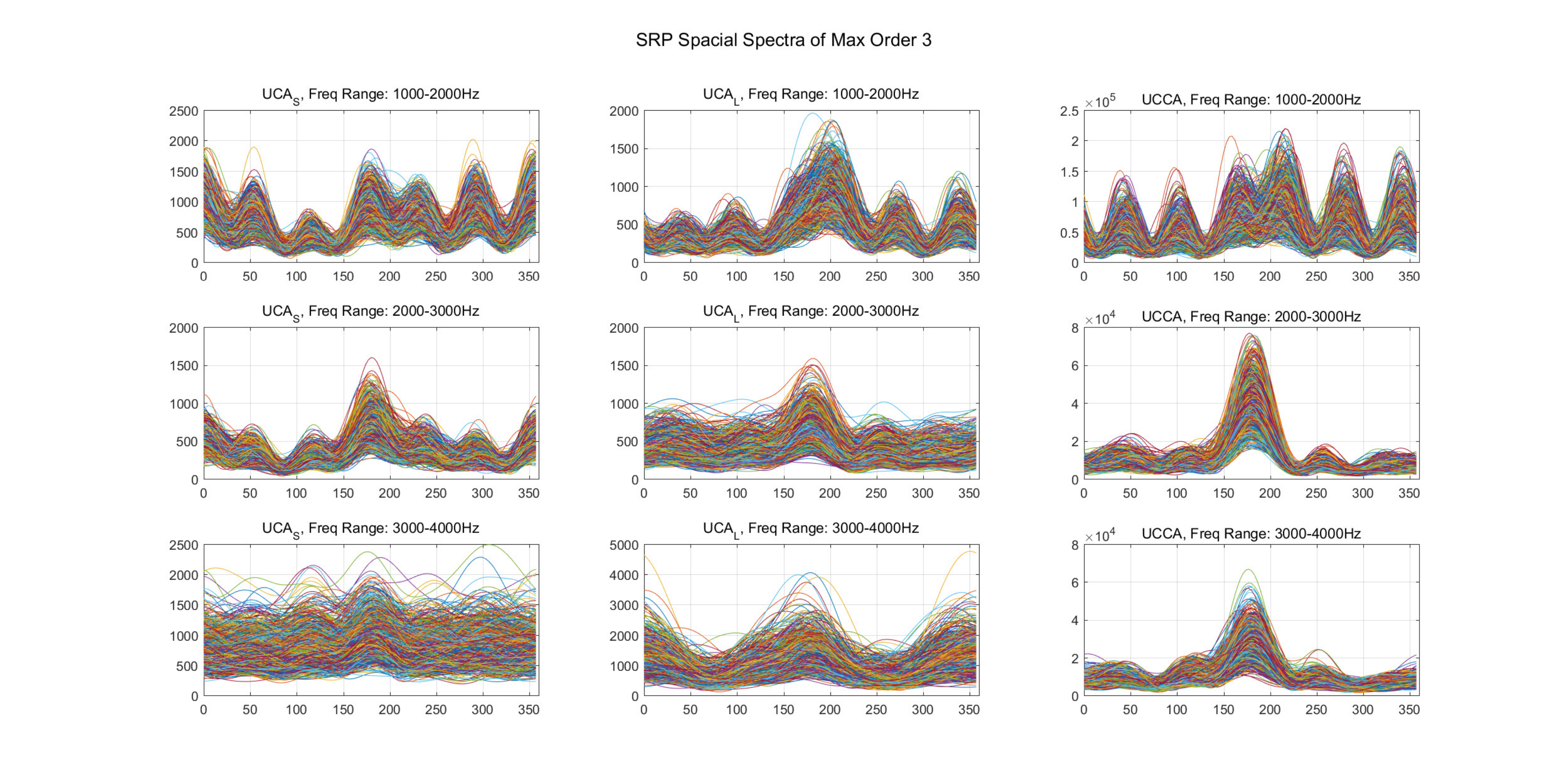}
\caption{SRP results of max order 1, 2, and 3.}
\end{figure}

By watching the spacial spectra shape under different conditions, it is evident to draw the following conclusions regarding the parameter settings:

\begin{enumerate}
    \item Generally, higher-order results exhibit a more obvious peak than the first order, which means a higher spatial resolution. This is in align with the circular harmonic patterns of different orders.
    \item Results from UCAs with max order 3 show poor estimating performance in low frequency band. This is because low values at low frequency band of high order Bessel functions (see Figure \ref{bessel}) lead to a low WGN of beamformers.
    \item By comparing the results of the different frequency range, the effects of zero points of Bessel functions  (see Figure \ref{bessel}) is clearly illustrated. Taking the smaller UCA as example, there is a zero-point between $3000Hz$ and $4000Hz$ in order 1, which not only makes the corresponding results invalid, but also affects the performance of order 2 and 3, since the higher-order results is an accumulation of all orders.
    \item UCCA shows a significant improvement compared with the other two UCAs under the same parameter settings. Statistical analysis shows higher accuracy of UCCA. The peak of UCCA results have a more clear peak too, which means a higher robustness in noisy and reverberation environment.
    \item Although in our approach we accumulate all results from selected frequency range on all orders, the estimating accuracy can be improved via an order-frequency choosing, discarding the "bad" results and combining the "good" results to produce a final better result.
\end{enumerate}

\subsection{Experiment Results}
Through the parameter analysis exhibited above, an optimized parameter setting can be derived, which is shown in Table 2.

\begin{table}[h]
\centering
\begin{tabular}{c c c}
\hline
\textbf{Max Order} & \textbf{Freq Range(kHz)} & \textbf{Average Length}\\
\hline
3 & [2,4] & 10 frames \\
\hline
\end{tabular}
\caption{Optimized parameter setting}
\end{table}

In addition, a time-domain average can be applied to acquire a more stable estimated result. In this case, we conduct a 10-frame average on the spacial spectrum, then find the maximum of the averaged special spectrum as the azimuth estimated. Thus, the averaged result is returned every $0.32s$.

The room environment and the experiment settings are exhibited in Figure 5. Inside the green circle is the UCCA used, while two speakers are set in $120^{\circ}, 1m$ and $240^{\circ}, 3m$ from the microphone array, respectively. It is understandable that the near speaker brings more direct sound and the far one causes more reflections, due to its location near the corner of the room.

\begin{figure}[htb]
    \centering
    \includegraphics[width=0.7\linewidth]{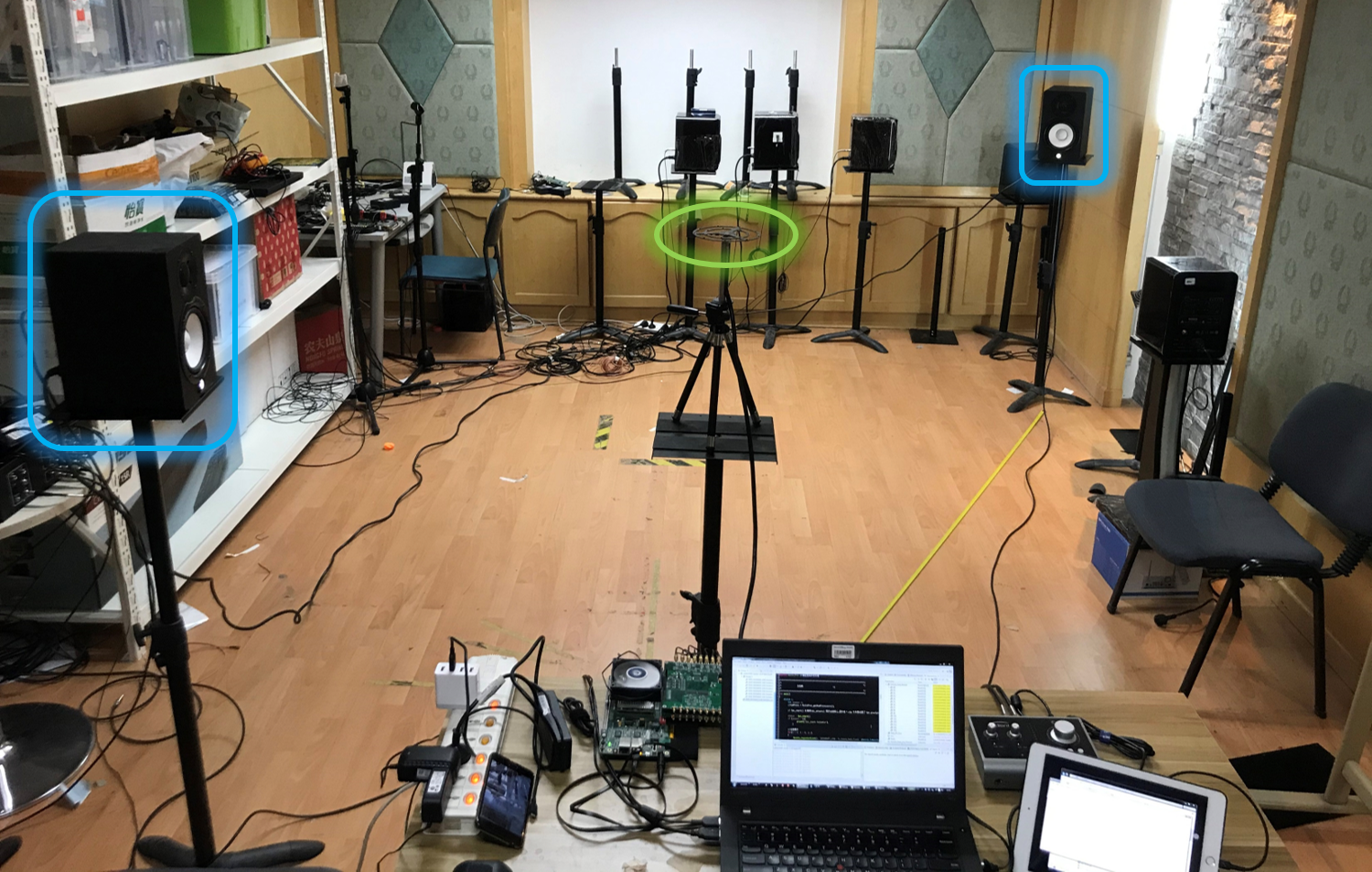}
    \caption{The room environment and the experiment settings.}
\end{figure}

We test the real-time performance of the prototype system at above mentioned two positions playing three types of sounds: noise, speech, and music. Noise audio is a Gaussian white noise clip generated by computer. The speech audio comes from ZeroSpeech 2019 dataset \cite{noauthor_download_nodate} . The music audio comes from Marsyas dataset \cite{Marsyas}. We collect 200 results of sound source localization of the system, presented in Figure 6.

\begin{figure}[!h]
    \centering
    \includegraphics[width=1\linewidth]{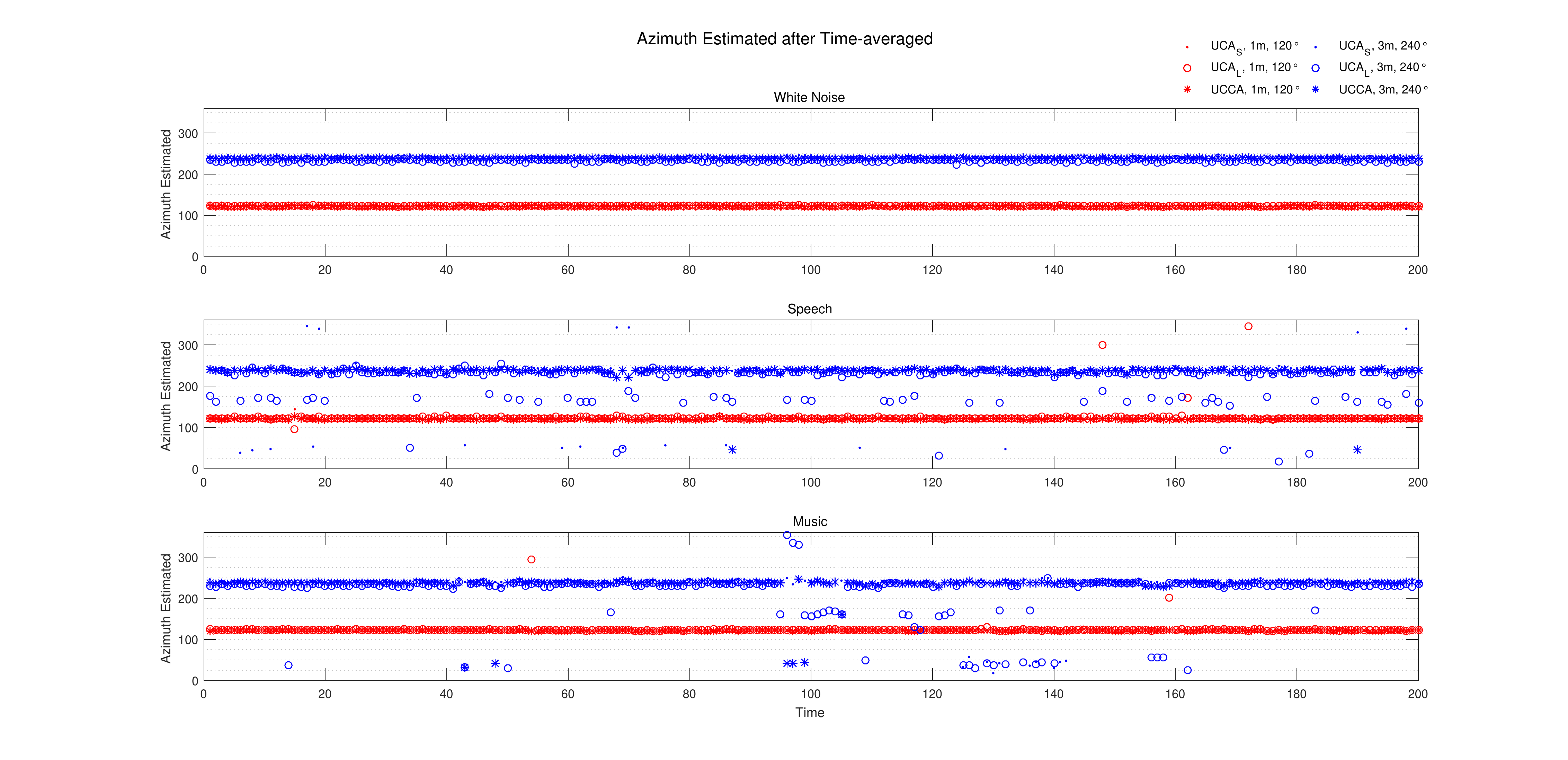}
    \caption{Real-time sound source localization results.}
\end{figure}

From the figure we can see that the system has a better performance towards stationary sounds like white noise. The dynamic of the sound source can affect the accuracy of the localization. On the other hand, the azimuth of the near source is estimated more accurately than the far source. It can be concluded that the reflections within a reverberation environment cause the failure of the localization.

We define a "success estimation" locating at $3^{\circ}$ range centered in its "real" position, then calculate the estimation success rate of above situations, shown as Table 3.

\begin{table}[h]
\centering
\begin{tabular}{c c c c c c c}
\hline
\multirow{2}*{\textbf{CMA}} & \multicolumn{3}{c}{\bm{$120^{\circ}, 1m$}} & \multicolumn{3}{c}{\bm{$240^{\circ}, 3m$}} \\
\textbf{} & \textbf{Noise} & \textbf{Speech} & \textbf{Music} & \textbf{Noise} & \textbf{Speech} & \textbf{Music}\\
\hline
$UCA_S$ & 0.765 & 0.985 & 0.990 & 0.730 & 0.250 & 0.245\\
$UCA_L$ & 0.970 & 0.965 & 0.985 & 0.555 & 0.170 & 0.100\\
$UCCA$ & 1.000 & 0.990 & 0.995 & 0.995 & 0.900 & 0.700\\
\hline
\end{tabular}
\caption{Success rate of sound source localization of CMAs.}
\end{table}

The success rate of UCCA is obviously higher than the UCAs, especially when the UCAs fails to locate the source in a strong reflecting environment.

\section{Conclusion}
\label{S:5}
In this paper, a real-time sound source localization system based on circular harmonics SRP method is implemented. The prototype system is capable of estimating azimuth of sound source in practical application. The performance under different parameter settings is discussed and an optimized set of parameters is derived. The system is verified in a real room with reverberation using playback of noise, speech, and music. Based on the prototype system, a performance comparison between UCAs and UCCA is conducted. The experiment results show that UCCA-based sound source localization is more accurate and robust, which still presents reasonable performance when UCA-based system is invalid due to the reflections in reverberation.

Further work will include refining the current system for better hardware performance, implementing better beamformer design methods, and developing programs that can selectively combine the information in various circular harmonic orders and frequency bins. Moreover, audio content analysis (ACA) techniques are proposed to be applied to select efficient information in the time-frequency domain by extracting audio features from the sound.


\end{document}